\documentclass[11pt]{article}
\usepackage{amsmath}
\usepackage{amsfonts}
\usepackage{amssymb}
\usepackage{graphicx}

\def\be\begin{equation}
 \def\ee{\end{equation}}
\def\bea{\begin{eqnarray}}
\def\eea{\end{eqnarray}}

\begin{document}
\begin{center}
\LARGE {\bf Brane  Cosmology for Vacuum and Cosmological Constant
Bulk }
\end{center}
\begin{center}
{\bf  Kh. Saaidi}\footnote{ksaaidi@uok.ac.ir},\\
{\bf  A. H. Mohammadi}\footnote{Abolhassan.Mohammadi@uok.ac.ir}\\

{\it Department of Physics, Faculty of Science, University of
Kurdistan,  Sanandaj, Iran}

\end{center}
 \vskip 1cm
\begin{center}
{\bf{Abstract}}
\end{center}
We consider the cosmology of a 3-brane universe, in a five
dimensional space time (Bulk). We present some solutions to the
five-dimensional Einstein equation, where a perfect fluid is
confined to the 3-brane. We investigate the evolution of brane
for two models of bulk. We choose ordinary and modified
polytropic gas in brane  and it is seen that these models have
some new features.
\\

\newpage

\section{Introductions}

Today, cosmology is an active field on physics. Its main goal is to describe
 the evolution of our universe from initial time to its present form. The
 mathematical description of cosmology is provided by the Einstein equations.
There has been a lot of activity in the domain of cosmology with
extra dimension. In
 the string theory the Einstein equations are part of the theory but the theory
 itself is consistent only in higher than four dimension.
  It has been suggested that {\cite{1}}, our
four-dimensional universe would be a 3-brane living in a
(4+d)-dimensional spacetime where d is the extra dimension. A
strong motivation of considering such models comes from strongly
coupled string theory {\cite{18}}. Indeed, Randall and Sundrum
{\cite{2},\cite{3}}, Geoberashvili {\cite{4}} for $d=1$,( reviving
the idea of {\cite{5}}), have shown that the extra dimension need
not even be compact. This result is in contrast with the main
investigation on the possibility of extra dimension which began
with Kaluza-Klein type theory (see e.g {\cite{6}}). The theory of
Kaluza-Klein explain that the extra dimensions are compactified
on a small enough radius to evade detection. Randall and Sundrum
have shown that these kind of models give back to standard
four-dimension gravity {\cite{7},\cite{8},\cite{9}}. For studying
the cosmological evolution of this model, somebody has assumed
that the density is localized on the brane, while the bulk energy
momentum tensor is only a negative cosmological constant
{\cite{7}}. Some modification of the cosmological evolution which
is due to the adding interaction in the gravitation sector, such
as Gauss-Banet term in the bulk {\cite{10}} and an induced gravity
term on the brane {\cite{11}}. Another modifications have
considered for the case which the bulk contains some matter
component in addition to the negative cosmological constant
{\cite{12},\cite{13},\cite{14}}.
 The cosmological evolution is studied for two system A and B. In
 system A, brane located at a fixed value of the
 fourth
 spatial coordinate, and the bulk is time-depended. And in the
 system B, the bulk is static and the brane is
 moving. In the case of a vacuum bulk, the brane evolution is
 discussed in two system of coordinate A {\cite{7},\cite{8}} and
 B {\cite{9},\cite{16}}.\\
 In the present work, our investigation is to solve the Einstein
 equation in the brane world scenario for a perfect
 fluid
 called polytropic gas on the brane for system A. It is well known
 that in standard cosmology a special kind of
  fluid is called
  polytropic gas {\cite{17}}.This kind of fluid has some very interesting
  properties, and hence,   in this paper we
  discussed the evolution of a homogeneous isotropic brane world
  filled with a polytropic and modified polytropic gas, when the
  bulk contain either vacuum or a cosmological
  constant.
  \par The paper is organized as follow: In section 2, we have
  introduced some basic equation that are useful
  for our investigation.
  In section 3, assuming a equation of state of polytropic gas for brane,
  then we have obtained energy density of
  brane with
  respect to scale factor of brane. In subsection 3.1, we have
  supposed a vacuum bulk, and it is shown that
  we have a deceleration universe for early
  time. In late time, it is shown that, we
  have a  universe with $\ddot{a_0}=0$. In subsection 3.2, we
  have assumed a cosmological constant bulk. So,
  it is shown that in early time we have similar condition
  as 3.1, but in late time, we can have an accelerate
  or decelerate universe. In section 4, we have chosen a modified
  polytropic for brane and obtained energy density of brane with
  respect to scale factor of brane. We consider two case for
  $\alpha$, and it is shown that the equation of state for
  $-1<\alpha<0$ is not a suitable case. Another case is
  $\alpha=1+\frac{1}{n}$.
  In subsection 4.1, we have a
  vacuum bulk for this kind of brane, and for early time we arrive
  at a decelerate universe; but for late time in contrary with the
  first section, we have an accelerate universe.
   In subsection 4.2, we have got a cosmological
  constant bulk model. It is shown that for early
  time the condition are similar to subsection 4.1, and we show that
  the cosmological constant, for this kind of brane, must be positive.

\section{Preliminary}
We suppose that the geometry of the five-dimensional bulk is characterized by a spacetime metric as:
\begin{equation}
ds^2=-n^2(t,y)dt^2+a^2(t,y)\delta_{ij}dx^idx^j+b^2(t,y)dy^2,
\end{equation}
where y is the fifth coordinate and $\delta_{ij}$ is a maximally
symmetric 3-dimensional metric. The Einstein equations in
5-dimensional is as follow:
\begin{equation}
G_{\alpha\beta}=R_{\alpha\beta}-\frac{1}{2}Rg_{\alpha\beta}={\kappa}^2T_{\alpha\beta},
\end{equation}
where $R=g^{\alpha\beta}R_{\alpha\beta}$, $R_{\alpha\beta}$ is
the 5-dimensional Ricci-tensor $\kappa$ is related to the
5-dimensional Newton's constant, $G_{(5)}$, and the 5-dimensional
reduced Plank mass $M_{(5)}$  is as:
\begin{equation}
{\kappa}^2=8 \pi G_{(5)}=M_{(5)}^{-3}.
\end{equation}
Now, for obtaining the solution of (2), we must specify the matter really. We classify the matter in two
distinct categories, i)matter  confined to our brane univers (3-brane), ii)matter distributed over the 5-dimensional
bulk. So that, the energy momentum tensor take the form:
\begin{equation}
T^{\alpha}_{\phantom{\alpha} \beta}=T^{\alpha}_{\phantom{\alpha}
\beta}|_{bu} +T^{\alpha}_{\phantom{\alpha} \beta}|_{br},
\end{equation}
Where $T^{\alpha}_{\phantom{\alpha} \beta}|_{bu}$ is the energy momentum of the bulk matter, and an explicit
form of that is
\begin{equation}
T^{\alpha}_{\phantom{\alpha} \beta}|_{bu}=diag\left(
-\rho_{bu}(t,y),p_{bu},p_{bu},p_{bu},p_{5bu} \right),
\end{equation}
the second term $T^{\alpha}_{\phantom{\alpha} \beta}|_{br}$ is related to the matter content in the brane $(y=0)$,
whereas we consider homogeneous and isotropic geometries inside the brane, one can write the energy-momentum tensor
in the brane as:
\begin{equation}
T^{\alpha}_{\phantom{\alpha}
\beta}|_{br}=\frac{\delta(y)}{b}diag\left(
-\rho_{br},p_{br},p_{br},p_{br},0 \right).
\end{equation}
So this assumption, clearly show that energy density,
$\rho_{br}$, and pressure, $p_{br}$, are independent of the
position inside the brane. It is seen that $T_{05}=0$, it means
that there is no flow of matter along the fifth dimension.
Therefore, using Eq.(1) and Eq.(2), one can arrive at
\begin{eqnarray}
G_{00}&=&3 \Bigg\{ \frac{\dot{a}}{a} \left( \frac{\dot{a}}{a}+\frac{\dot{b}}{b} \right) - \frac{n^2}{b^2}
\left( \frac{a''}{a} + \frac{a'}{a} \left( \frac{a'}{a} - \frac{b'}{b} \right) \right) + k \frac{n^2}{a^2} \Bigg\}, \\
G_{ij}&=&\frac{a^2}{b^2}\gamma_{ij}\Bigg\{ \frac{a'}{a} \left( \frac{a'}{a}+2\frac{n'}{n} \right) - \frac{b'}{b}
 \left( \frac{n'}{n}+2\frac{a'}{a} \right) +2\frac{a''}{a}+\frac{n''}{n} \Bigg\} {} \nonumber \\
 & & + \frac{a^2}{n^2}\gamma_{ij} \Bigg\{ \frac{\dot{a}}{a} \left( -\frac{\dot{a}}{a}+2\frac{\dot{n}}{n} \right)
 - 2\frac{\ddot{a}}{a} + \frac{\dot{b}}{b}\left( -2\frac{\dot{a}}{a}+\frac{\dot{n}}{n}
 \right),
 - \frac{\ddot{b}}{b} \Bigg\} - k\gamma_{ij}\\
G_{05}&=&3\left( \frac{n'}{n}\frac{\dot{a}}{a} + \frac{a'}{a}\frac{\dot{b}}{b} - \frac{\dot{a'}}{a} \right),\\
G_{55}&=&3\Bigg\{ \frac{a'}{a}\left( \frac{a'}{a}+\frac{n'}{n}
\right) - \frac{b^2}{n^2} \left( \frac{\dot{a}}{a} \left(
\frac{\dot{a}}{a}-\frac{\dot{n}}{n} \right) \right) -
k\frac{b^2}{a^2} \Bigg\},
\end{eqnarray}
where $\dot{\chi}=\frac{d\chi}{dt}$ and $\chi'=\frac{d\chi}{dy}$.
In order to have a well defined geometry, the metric tensor must
be continuous across the brane (i.e $y=0$) while its derivative
with respect to y can be discontinuous on the brane. As a result,
the Dirac delta function will appear in the second-order
derivative of metric coefficient with respect to y. So according
to [7], we can obtain
\begin{equation}
\frac{[a']}{b_0a_0}=-\frac{\kappa^2}{3}\rho_{br},
\end{equation}
\begin{equation}
\frac{[n']}{b_0n_0}=\frac{\kappa^2}{3}(3p_{br}+2\rho_{br}),
\end{equation}
where the subscript 0 in the scale factor indicate their values
on the brane. We take the jump of Eq.(9) and using
Eqs.(11,12), we can obtain
\begin{equation}
\dot{\rho}_{br}+3\frac{\dot{a}_0}{a_0}(\rho_{br}+p_{br})=0,
\end{equation}
then, the average value of Eq.(10) for $y\longrightarrow 0^+$ and
$y\longrightarrow 0^-$ and impose the $Z_2$-symmetry, we arrive
at {\cite{7}}
\begin{equation}
\frac{\ddot{a}_0}{a_0}+\frac{\dot{a}_0^2}{a_0^2}=-\frac{\kappa^4}{36}
\rho_{br}(\rho_{br}+3p_{br}) -\frac{\kappa^2}{3b_0^2}
p_{5bu}-\frac{k}{a_0^2}
\end{equation}
By using a suitable time transformation, we can choose $n_0=1$.
Eq.(14) is known as generalize Friedmann type equation on the
brane. It has a basic difference from the ordinary Friedmann
equation of standard cosmology. In fact, Eq.(14) H depends
quadratically on the brane energy density, in contract with the
usual linear
dependence in a ordinary Friedmann equation.\\
We can obtain the 5D-field equations (7) and (10) in the compact
form as {\cite{7}}:
\begin{eqnarray}
\psi'&=&-\frac{2}{3}a'a^3\kappa^2\rho_{bu},
\end{eqnarray}
\begin{eqnarray}
\dot{\psi}&= &\frac{2}{3}\dot{a}a^3\kappa^2p_{5bu},
\end{eqnarray}
where
\begin{equation}
\psi\equiv\frac{(a'a)^2}{b^2}-\frac{(\dot{a}a)^2}{n^2}-ka^2.
\end{equation}
By equating the y-derivative of Eq.(16) with the time-derivative
of Eq.(15), we have
\begin{equation}
a'\dot{\rho_{bu}}+\dot{a}p'_{5bu}+(\rho_{bu}+p_{5bu})\left(
\dot{a'}+3\frac{\dot{a}a'}{a} \right)=0,
\end{equation}
and, the constraint $\nabla_\alpha G^{\alpha0}=0$, gives
\begin{equation}
\dot{\rho_{bu}}+3\frac{\dot{a}}{a}(\rho_{bu}+p_{bu})+\frac{\dot{b}}{b}(\rho_{bu}+p_{5bu})=0.
\end{equation}
Using the results of junction condition (11) and imposing the
$Z_2$-symmetry and also taking an mean value of equation (17) for
$y\longrightarrow 0^+$ and $y\longrightarrow 0^-$, one can obtain
the generalize Friedmann equaton  on the brane as
\begin{equation}
\frac{\dot{a}_0^2}{a_0^2}=\frac{\kappa^4}{36}\rho_{br}^2-\frac{\psi_0(t)}{a_0^4}-\frac{k}{a_0^2}.
\end{equation}

\section{Polytropic Brane Model}
We now consider a polytrop gas model of DE {\cite{17}} which
satisfies the equation of state
\begin{equation}
p_{br}=A\rho^{1+\frac{1}{n}}_{br},
\end{equation}
where A is a constant and n is the polytropic index{\cite{18}}.
From the energy conservation equation (13),
 the expression for matter density is given by
\begin{equation}
\rho_{br}={\left( \frac{1}{\rho_1a_0^{\frac{3}{n}}-A} \right)}^n,
\end{equation}
where $\rho_1$ is an arbitrary integration constant. We clearly
see that if $A$ be negative, $a_0$ can lead to zero. But if $A$
be positive, $a_0$ can not lead to zero, because with pass of
time  from zero, the denominator of Eq.(22) decrease and
$\rho_{br}$ increase. Then in special time the energy density
goes to infinite. It means with pass of time, energy density
increase and physically it is not true. So for positive value of
$A$ the model does not have a singularity, and we have a minimum
value for brane scale factor in early time i.e $a_0$ should be
 larger than $ \left( \frac{A}{\rho_1}
\right)^{\frac{n}{3}} $  .

\subsection{Vacuum Bulk With Polytropic Gas in Brane }
Whereas there is no matter in the Bulk, we have
$T^{\alpha}_{\phantom{\alpha} \beta}|_{bu}=0$. Therefore,
Eqs.(15) and (16) show that $\psi=D$ should be constant.
Then, the generalized Friedmann equation is as
\begin{equation}
\left(
\frac{\dot{a}_0}{a_0}\right)^2=\frac{\kappa^4}{36}\rho_{br}^2 -
\frac{D}{a_0^4}-\frac{k}{a_0^2},
\end{equation}
and from Eq.(14) we have
\begin{equation}
\frac{\ddot{a}_0}{a_0}+\frac{\dot{a}_0^2}{a_0^2}=-\frac{\kappa^4}{36} \rho_{br}(\rho_{br}+3p_{br})
- \frac{k}{a_0^2}.
\end{equation}
By using these results we arrive at:
\begin{equation}
\dot{a_0}^2=\frac{\kappa^4}{36}{\left( \frac{1}{\rho_1a_0^{\frac{3}{n}}-A} \right)}^{2n} a_0^2
    - \frac{D}{a_0^2}-k,
\end{equation}
and
\begin{eqnarray}
\ddot{a_0}&=&-\frac{\kappa^4}{18}{\left(
\frac{1}{\rho_1a_0^{\frac{3}{n}}-A} \right)}^{2n} a_0 -
\frac{\kappa^4}{12}A{\left( \frac{1}{\rho_1a_0^{\frac{3}{n}}-A}
\right)}^{2n+1} a_0  \nonumber \\
 & & +\frac{D}{a_0^3}.
\end{eqnarray}
Here, we continue our investigation for various value of $A$:

\begin{itemize}
\item{\bf For $A<0$}\\
From (25) and (26), we realize when $a_0\longrightarrow 0$, $\dot{a_0}^2$ lead to
$-\frac{D}{a_0^2}$. Also $\dot{a_0}^2$ must be positive, as a result
\begin{equation}
D<0.
\end{equation}
In this era $\ddot{a_0}\longrightarrow  \frac{D}{a_0^3}$.
Because $D$ is a negative constant,
we have a decelerate universe in early time.\\

In late time (i.e $a_0\longrightarrow \infty$),
$\dot{a_0}^2\longrightarrow -k$ and
 $\ddot{a_0}\longrightarrow 0$, too. Also, $k$ can be $0$ or $-1$.

\item{\bf For $A>0$ }\\
 In this case for investigation of universe in early time we should lead $a_0$ toward to
 $\left( \frac{A}{\rho_1} \right)^{\frac{n}{3}}$. So, as a result $\dot{a_0}^2\longrightarrow +\infty$
  and $\ddot{a_0}\longrightarrow -\infty$. It shows that we have a decelerate universe.
  For late time when we lead $a_0$ toward to infinity, $\dot{a_0}^2\longrightarrow -k$ and
  $\ddot{a_0}\longrightarrow 0$; like to the previous case.
\end{itemize}

\subsection{Cosmological Constant Bulk With Polytropic Gas in \\
Brane} In this case, we choose equation of state as
$p_{bu}=p_{5bu}=-\rho_{bu}$ in the bulk. From conservation
relation (18) and (19) we arrive at this result that,
$\rho_{bu}=\Lambda$ must be a constant. According to Eqs.(15,16),
we have
\begin{equation}
\psi(t,y)=-\frac{\kappa^2}{6}\Lambda a^4+D,
\end{equation}
D is a integration constant.
Now, with substituting this result in evolution equations of brane, we obtain
\begin{equation}
\dot{a_0}^2=\frac{\kappa^4}{36}\left(
\frac{1}{\rho_1a_0^{\frac{3}{n}}-A} \right)^{2n}a_0^2
+\frac{\kappa^2}{6}\Lambda a_0^2-\frac{D}{a_0^2}-k,
\end{equation}
and
\begin{eqnarray}
\ddot{a_0}&=&-\frac{\kappa^4}{18} \left( \frac{1}{\rho_1a_0^{\frac{3}{n}}-A} \right)^{2n} a_0
- \frac{\kappa^2}{12}A \left( \frac{1}{\rho_1a_0^{\frac{3}{n}}-A} \right)^{2n+1}a_0 \nonumber \\
 & &- \frac{\kappa^2}{3}(\frac{1}{2}-\frac{1}{b_0^2})\Lambda a_0 +\frac{D}{a_0^3}.
\end{eqnarray}
In the follow, we investigate evolution of universe for negative
and positive values of $A$:

\begin{itemize}
\item{\bf For $A<0$:}\\
In this case, we can lead $a_0$ toward to zero, so when $a_0\longrightarrow 0$,
$\dot{a_0}\longrightarrow -\frac{D}{a_0^2}$. For having a positive $\dot{a_0}^2$,
$D$ must be negative; namely
\begin{equation}
D<0.
\end{equation}
Also, in this time, $\ddot{a_0}\longrightarrow
\frac{D}{a_0^3}$. This situation is like to the previous subsection, and
 we again have a decelerate universe in early time.\\
 In late time (when $a_0\longrightarrow \infty$), $\dot{a_0}^2\longrightarrow \frac{\kappa^2}{6}\Lambda a_0^2$.
 Therefore, because $\dot{a_0}^2$ is positive, so $\Lambda$ must be positive ($\Lambda >0$). Also, in this era
 $\ddot{a_0}\longrightarrow - \frac{\kappa^2}{3}(\frac{1}{2}-\frac{1}{b_0^2})\Lambda a_0$. It means if $b_0^2>2$,
 we have a deceleration and if $b_0^2<2$ we have an acceleration. Note that if $b_0^2=2$, $\ddot{a_0}$ is equal to zero,
  and we do not have any deceleration or acceleration universe.

\item{\bf For $A>0$:}\\
In this case, $a_0$ tends to $\left( \frac{A}{\rho_1}
\right)^{\frac{n}{3}}$ for investigation of universe on early
time. As $a_0\longrightarrow \left( \frac{A}{\rho_1}
\right)^{\frac{n}{3}}$; $\dot{a_0}^2\longrightarrow +\infty$ and
$\ddot{a_0}\longrightarrow -\infty$. So, it is shown that we have
decelerate universe. For late time, when $a_0\longrightarrow
\infty$, $\dot{a_0}^2\longrightarrow \frac{\kappa^2}{6}\Lambda
a_0^2$. It is clearly seen that, is $\Lambda$ be positive
$\dot{a_0}^2$
 will become a positive quantity. Also $\ddot{a_0}\longrightarrow -
 \frac{\kappa^2}{3}(\frac{1}{2}-\frac{1}{b_0^2})\Lambda a_0$, because of $\Lambda >0$,
  we have a decelerate universe for $b_0^2>2$, and an accelerate universe for $b_0^2<2$.
  And if $b_0^2=2$ there is not any deceleration or acceleration universe.

\end{itemize}

\section{Modified Polytropic Brane Model}
In this section we get another equation of state for brane as:
\begin{equation}
p_{br}=-\rho_{br}+A\rho_{br}^{\alpha},
\end{equation}
where A an $\alpha$ are constant. With the help of brane
conservation equation, Eq.(13), one can obtain
\begin{equation}
\rho_{br}^{1-\alpha}=\ln \left( \frac{1}{\rho_1a_0^{3A(1-\alpha)}}
\right).
\end{equation}
$\rho_1$ is a integration constant. IN the follow, we discuss
about two case for $\alpha$.

\begin{itemize}

\item{\bf If $-1\leq\alpha<0$.}\\
In this case equation (32) shows  the equation of state of
modified chaplygin gas. According to Eq.(33), we arrive at
\begin{equation}
\rho_{br}= \Big( -3A(1-\alpha) \ln(\rho'_1 a_0)
\Big)^{\frac{1}{1-\alpha}}.
\end{equation}
If $A$ be a positive constant $\rho_{br}$ does not become a
physical. So, we take $A$ as a negative constant. With a little
attention , we realized that, $a_0$ can not be smaller than
$\left( \frac{1}{\rho'_1} \right)$. It means we have a minimum
value for $a_0$ in initial time. Here, it is clearly seen that the
energy density tends to infinite for late time, and this is
incompatible with experiment. Then, this case is not a suitable
equation of state; also, for $\alpha<1$ we have this situation.
So, the equation of state which are combination of cosmological
constant and every term of $A\rho^{\alpha}$ (where $\alpha<1$)
are not a good equation of state. For instance, the equation of
state of chaplygin gas and generalized chaplygin gas can not be
with cosmological constant considered.

\item{\bf If $\alpha=1+\frac{1}{n}$.}\\
In this case equation (32) shows the equation of state of
modified polytropic gas, also in here we take $A$ as a positive
constant. So, from (33) we have
\begin{equation}
\rho_{br}= \left( \frac{n}{3A  \ln(\rho'_1 a_0)} \right)^n.
\end{equation}
From this relation, it is clearly seen that $a_0$ can not be
smaller than $\left( \frac{1}{\rho'_1} \right)$, so at initial
time we have a minimum size for $a_0$, and in that time $a_0$
should be larger than $\left( \frac{1}{\rho'_1} \right)$. Then
for investigation of universe in early time $a_0$ tends to one.

\end{itemize}

At the continuance we investigate the evolution of universe for
two model of bulk.

\subsection{Vacuum Bulk With Modified Polytropic Gas in Brane}
Here we assume vacuum for bulk. Because we have no matter in
the bulk, we can write $T^{\alpha}_{\phantom{\alpha}
\beta}|_{bu}=0$. So according to (15) and (16), $\psi=D$ is
constant. Now, we obtain the evolution of universe as follow.
Substituting (35) in (14) and (20),

 it is easily obtained
\begin{equation}
\dot{a_0}^2=\frac{\kappa^4}{36} \left( \frac{n}{3A \ln(\rho'_1
a_0)} \right)^{2n} a_0^2 - \frac{D}{a_0^2}-k,
\end{equation}
also
\begin{eqnarray}
\ddot{a_0}&=&\frac{\kappa^4}{36} \left( \frac{n}{3A \ln(\rho'_1
a_0)} \right)^{2n} a_0 - \frac{\kappa^4}{12} A \left(
\frac{n}{3A  \ln(\rho'_1 a_0)} \right)^{2n+1} a_0 \nonumber \\
 & &+\frac{D}{a_0^3}.
\end{eqnarray}
From (36) and (37), we consider th state of our universe. In early
time $a_0$ tends to $\left( \frac{1}{\rho'_1} \right)$, so
$\dot{a_0}^2\longrightarrow +\infty$ and
 $\ddot{a_0}\longrightarrow -\infty$ $$ \ddot{a_0}\longrightarrow \frac{\kappa^4}{12}\left( \frac{n}{3A
 \ln(\rho'_1 a_0)} \right)^{2n} a_0 \Big( \frac{1}{3}-\frac{n}{3
\ln(\rho'_1 a_0)} \Big)$$; it means that, we have a
decelerate universe in early time.\\
In late time when $a_0$ tends to infinite,
$\dot{a_0}^2\longrightarrow +\infty$, also
$\ddot{a_0}\longrightarrow +\infty$. So, we have an accelerate
universe in late time.

\subsection{Cosmological Constant Bulk With Modified Polytropic Gas in Brane}
In this case, we impose $p_{bu}=p_{5bu}=-\rho_{bu}$ as equation
of state of bulk. Using (18) and(19) we arrive to
 this result that $\rho_{bu}=\Lambda$ is constant. Eqs.(15) and (16) and
 this result help us to write $\psi$ as:
\begin{equation}
\psi(t,y)=-\frac{\kappa^2}{6}\Lambda a^4+D.
\end{equation}
In the continuance, like to the previous subsection we
investigate the evolution of universe for some specific value of
$\alpha$.

The evolution equations are as
\begin{equation}
\dot{a_0}^2=\frac{\kappa^4}{36} \left( \frac{n}{3A \ln(\rho'_1
a_0)} \right)^{2n} a_0^2 + \frac{\kappa^2}{6}\Lambda a_0^2 -
\frac{D}{a_0^2}-k,
\end{equation}
and
\begin{eqnarray}
\ddot{a_0}&=&\frac{\kappa^4}{36} \left( \frac{n}{3A \ln(\rho'_1
a_0)} \right)^{2n} a_0 - \frac{\kappa^4}{12} A \left(
\frac{n}{3A  \ln(\rho'_1 a_0)} \right)^{2n+1} a_0 \nonumber \\
 & &+ \frac{\kappa^2}{6}\Lambda a_0 +\frac{D}{a_0^3}.
\end{eqnarray}
From(39) and (40), as $a_0\longrightarrow \left( \frac{1}{\rho'_1}
\right)$, then $\dot{a_0}^2\longrightarrow +\infty$ and
$\ddot{a_0}\longrightarrow -\infty$. It means we have a
decelerate universe in early time.\\
As $a_0\longrightarrow \infty$, $\dot{a_0}^2\longrightarrow
\frac{\kappa^2}{6}\Lambda a_0^2$, then $\Lambda$ must be positive
to give a positive value for $\dot{a_0}^2$. Also,
$\ddot{a_0}\longrightarrow \frac{\kappa^2}{6}\Lambda a_0$;
namely, we have an accelerate universe in late time.

\section{conclusion}
It has shown that for polytropic model of barne with $A<0$, we
have a limited value for brane energy density at initial time,
and for $A>0$ we have a nonzero value for brane scale factor. It
is shown that, for vacuum bulk model we have a decelerate
universe; Also in late time for both positive and negative value
of $A$, we have $\ddot{a_0}=0$. In the cosmological bulk model,
the condition of universe in early time is like to the vacuum bulk
model; however, in late time dependence on the
 value of $b_0^2$, we can have an accelerate universe. \\
In the second model of brane we choose two case for $\alpha$. For
$-1\leq\alpha<0$, it is shown that the constant in equation of
state should be negative, also it is realized that this case is
not a suitable model. When $\alpha=1+\frac{1}{n}$, it has shown
that $a_0$ in initial time has a minimum value. we find that for
two model of bulk there is a decelerate universe in early time
and an accelerate universe in late time; also we arrive at this
result that, the cosmological constant of bulk, in this model,
must be positive.


\begin{thebibliography}{99}


\bibitem{1} N. Arkani-Hamed, S. Dimopoulos, G. Dvali, Phys.Lett. B {\bf 429}, 263
(1998).


\bibitem{2} L. Randall, R. Sundrum, Phys. Rev. Lett. {\bf 83}, 3370 (1999).

\bibitem{3} L. Randall, R. Sundrum, Phys. Rev. Lett. {\bf 83}, 4690 (1999).

\bibitem{4} M. Gogberashvili, hep-ph/9812365; hep-ph/9904383.

\bibitem{5} V. A. Rubakov, M. E. Shaposhnikov, Phys. Lett. B {\bf 125}, 139 (1983);\\
M. Visser, Phys. Lett. B {\bf 159}, 22 (1985); \\
E. J. Squires, Phys. Lett. B {\bf 167}, 286 (1986);\\
G. W. Gibbons, D. L. Wiltshire, Nucl. Phys. B {\bf 287}, 717
(1987).
\bibitem{6} D. Bailin, A. Love, Rep. Prog. Phys. 50, 1087 (1987)



\bibitem{7}P. Binetruy, C. Deffayet, U. Ellwanger, D. Langlois,
Phys. Lett. B {\bf 477}, 285 (2000);
 Nucl. Phys. B {\bf 565}, 269 (2000).\\
P. Binetruy, C. Deffayet, U. Ellwanger, D. Langlois, Phys. Lett. B
{\bf 477}, 285 (2000).

\bibitem{8} C. Csaki, M. Graesser, C. F. Kolda and J. Terning, Phys. Lett. B {\bf 462}, 34
(1999);\\

J. M. Cline, C. Grojean and G. Servant, Phys. Rev.
Lett. {\bf 83}, 4245 (1999);\\
A. Padilla, Braneworld Cosmology and Holography, Ph.D. thesis,
(2002) University of Durham, England [arXiv:hep-th/0210217].

\bibitem{9}P. Kraus, JHEP {\bf 9912}, 011 (1999).

\bibitem{10} N. Deruelle and T. Dolezel, Phys. Rev. D {\bf 62}, 103502 (2000);\\
B. Abdesselam and N. Mohammedi, Phys. Rev. D {\bf 65},084018
(2002);\\
C. Germani and C. F. Sopuerta, Phys. Rev. Lett. {\bf 88}, 231101
(2002).

\bibitem{11}H. Collins and B. Holdom, Phys. Rev. D {\bf 62}, 105009 (2000);\\
E. Kiritsis, N. Tetradis and T. N. Tomaras, JHEP {\bf 0203}, 019
(2002).

\bibitem{12} C. van de Bruck, M. Dorca, C. J. A. Martins and M. Parry, Phys.
Lett. B {\bf 495}, 183 (2000);\\
T. N. Tomaras, arXiv:hep-th/0404142;\\
E. Kiritsis, Fortsch. Phys. {\bf 52}, 568 (2004).

\bibitem{13} E. Kiritsis, G. Kofinas, N. Tetradis, T. N. Tomaras and V. Zarikas,
JHEP {\bf 0302}, 35 (2003);\\
 N. Tetradis, Phys. Lett. B {\bf 569}, 1 (2003).

\bibitem{14} P. S. Apostolopoulos and N. Tetradis, Class. Quant.
Grav. {\bf 21}, 4781 (2004);\\
N. Tetradis, Class. Quant. Grav. {\bf 21}, 5221 (2004).

\bibitem{15} A. Hebecker and J. March-Russell, Nucl. Phys.
B {\bf 608}, 375 (2001).

\bibitem{16} D. Birmingham, Class. Quant. Grav. {\bf 16}, 1197 (1999).


\bibitem{16} K. Karami, S. Gaffari, J. Fehri, Eur. Phys. J. C (2009), 64, 85,
and arXiv:0911.4915v1 [gr-qc].

\bibitem{17}J. Christensen-Dalsgaard,
Lecture Notes on Stellar Structure and Evolution, 6th edn. (Aarhus
University Press, Aarhus, 2004)

\bibitem{18} P. Ho¡rava and E. Witten, Nucl. Phys. B {\bf 460}, (1996) 506, Nucl. Phys. B {\bf
475}, 94 (1996);\\
I. Antoniadis, N. Arkani-Hamed, S. Dimopoulos, G. Dvali, Phys.
Lett. B {\bf 436}, 257 (1998) .

\end{thebibliography}
\end{document}